\documentclass{emulateapj}
\usepackage{latexsym}
\usepackage{dcolumn}
\usepackage{bm}
\input{epsf}
\newcommand{\be}{\begin{equation}}
\newcommand{\ee}{\end{equation}}
\newcommand{\ba}{\begin{eqnarray}}
\newcommand{\ea}{\end{eqnarray}}
\newcommand{\bfi}{\begin{figure}
\epsfxsize=9cm
\epsffile}
\newcommand{\efi}{\end{figure}}

\begin{document}
\title{Isolating the decay rate of cosmological gravitational potential}
\author{Pengjie Zhang}
\email{pjzhang@shao.ac.cn}
\affil{Shanghai Astronomical Observatory, Chinese Academy of
  Science, 80 Nandan Road, Shanghai, China, 200030}
\affil{Joint Institute for Galaxy and Cosmology (JOINGC) of
SHAO and USTC}
\begin{abstract}
The decay rate of cosmological gravitational potential measures the
deviation from Einstein-de Sitter universe and can put 
strong constraints on the nature of dark energy and gravity. Usual
method to measure this decay rate is through the integrated Sachs-Wolfe (ISW) effect-large scale
structure (LSS) cross correlation. However, the interpretation of the
measured correlation signal is complicated by the galaxy bias and matter
power spectrum. This could bias and/or degrade its constraints to the nature of dark
energy and gravity. But, combining the lensing-LSS cross
correlation measurements, the decay rate of gravitational
potential can be isolated. For any given narrow redshift bin of LSS, the ratio
of the two cross correlations directly measures $[d\ln D_{\phi}/d\ln
  a]H(z)/W(\chi,\chi_s)$, where $D_{\phi}$ is the linear growth  factor
of the gravitational potential, $H$ is the Hubble constant at redshift
$z$, $W(\chi,\chi_s)$ is the lensing kernel and  $\chi$ and $\chi_s$
are the comoving 
angular diameter distance to lens and source, respectively. This
method is optimal in the sense that (1) the measured quantity is
essentially free of systematic errors and is only limited by cosmic
variance and (2) the measured quantity only depends on several
cosmological parameters and can be predicted from first principles
unambiguously.  Though fundamentally limited by inevitably large
cosmic variance associated with the ISW measurements, it can still put useful
independent constraints on  the  
amount of dark energy and its equation 
of state. It can also  provide a powerful test of modified gravity and can
distinguish the Dvali-Gabadadze-Porrati model from $\Lambda$CDM 
 at $>2.5\sigma$ confidence level. 
\end{abstract}
\ \maketitle

\section{Introduction}
The decay rate of gravitational potential is a
powerful probe of dark energy \citep{ISWcross} and the nature of
gravity \citep{Lue04a,Lue04b,Zhang05}.  Its direct observational
consequence is the integrated Sachs-Wolfe
(ISW) effect \citep{ISW}, one class of secondary CMB temperature
fluctuations. However, since the signal is overwhelmed by 
primary cosmic microwave background (CMB),
it has to be  measured by cross correlating the large
scale structure (LSS) \citep{ISWcross,Seljak99b}, or from CMB polarization induced by
intra-cluster electron scattering \citep{Cooray04}.  Progresses in precision CMB measurements and
large scale galaxy surveys have enabled detections of the ISW-LSS
cross correlation and confirmed the decay of cosmological 
gravitational potential \citep{Fosalba03,Scranton03,Afshordi04,Boughn04,Fosalba04,Nolta04,Vielva04,Padmanabhan05}.  Along with the flatness of the
universe constrained from the CMB \citep{Spergel03},  detections of the
gravitational potential decay have already provided strong evidence of the
existence of dark energy \citep{Corasaniti05} and constraints to the
dark energy equation of state can be further improved by future
observations \citep{Pogosian05}. However, the
strength of the cross 
correlation signal depends not only on the dark energy density
$\Omega_{\rm DE}$ and its equation of state parameter 
$w_{\rm DE}$, but also  on the matter power spectrum and evolving galaxy
bias\footnote{The cross correlation signal is proportional to $r$, the
  cross correlation coefficient between galaxy 
  overdensity and the gravitational potential.  In parameter fitting
  combining galaxy power spectrum measurement,  $r=1$ at relevant
  scales is implicitly 
  assumed. But stochasticity could cause $r$ to deviate from unity. So
  in principle, $r$ should also be 
  treated as  a free parameter 
  to be marginalized.}, which can not be predicted from first principles. These nuisance
parameters and modeling uncertainties could  degrade the power
of ISW-LSS  
cross correlation to constrain dark energy. However, as proposed
in this paper, by the aid of gravitational lensing-LSS cross
correlation, the evolution of gravitational potential can be
isolated. 

The ISW effect and lensing-LSS probe the same 3D gravitational
potential $\phi$, with different prefactors. Since 
galaxy redshifts are observable, one can correlate ISW or lensing with
galaxies in a narrow redshift bin. For such
narrow redshift  bin, the ISW-galaxies cross
correlation measures  $\dot{\phi}$-$\delta_g$ cross correlation,
while the lensing-galaxies cross correlation 
  measures $\phi$-$\delta_{g}$ cross correlation. Here,
  $\delta_{g}$ is the over-density of galaxies. The ratio of the two correlations in
  the same redshift bin at the same scale then measures
  $\dot{\phi}/\phi$, with  
  prefactor which only depends on the geometry of the Universe, but
  not  galaxy bias nor matter power spectrum. Such measurement
  involves essentially no assumptions and least amount of unknown
  cosmological  parameters, so it can put robust constraints on cosmology
  and has the power to distinguish dark energy from modified
  gravity. Furthermore, since these two cross
  correlation measurements are correlated, errors in the denominator 
  and numerator partly cancel. The S/N of the measured ratio is
  slightly better than  the ISW-LSS measurement. 
 Since $\dot{\phi}/\phi$ of each redshift can be recovered, one
 obtains stronger cosmological constraints than that of the
  ISW-lensing measurement \citep{Seljak99b}. 

\section{Isolating the decay rate of gravitational potential}
 Time variation in the cosmological gravitational potential causes
temperature fluctuations in CMB \citep{ISW} 
\be
\frac{\Delta T}{T_{\rm CMB}}=\int [\dot{\phi}-\dot{\psi}]ad\chi \ ,
\ee
where $\phi$ and $\psi$ are two gravitational potentials in the
Newtonian gauge. In dark energy models, at late time, there is no
anisotropic stress and $\phi=-\psi$.  The ISW-LSS cross correlation power
spectrum is given by \citep{Peebles73}
\be
\label{eqn:Ig}
C_{Ig}=4\pi
\int
\Delta^2_{\phi\delta}(k,z=0)\frac{dk}{k}A_I(k,l)A_g(k,l)\ .
\ee
Here, galaxies at $\bar{z}-\Delta z/2<z<\bar{z}+\Delta z/2$ have
  been adopted to be tracers of the LSS. $A_g= 
\int_{\chi_1}^{\chi_2} j_l(k\chi)n(\chi)D_{g}d\chi$ and $A_I\equiv
2\int_0^{\chi_{\rm CMB}} 
  j_l(k\chi)\dot{D}_{\phi}ad\chi$, where $\chi_1,\chi_2,\chi_{\rm
    CMB}$ are the comoving angular diameter distance to
  $\bar{z}-\Delta z/2$, $\bar{z}+\Delta z/2$ and the last
    scattering surface, respectively. $D_g$ and $D_{\phi}$ are
  the linear growth factors of galaxy over-density and gravitational
  potential $\phi$, respectively.  $n_g(\chi)$ is the number of galaxies per
  distance interval. $\Delta^2_{\phi\delta}$ is the cross correlation
  power spectrum variance between $\phi$ and galaxy over-density
  $\delta$. $j_l$ is the spherical Bessel function. This equation
  assumes linear evolution in $\phi$ and galaxy over-density, which
  should be valid at large scales where almost all ISW signal comes
  from. But it does  not assume the galaxy bias to be scale
  independent. Eq. \ref{eqn:Ig} and \ref{eqn:phig} (please refer to
  the appendix for the derivation) do not require the 
  small angle approximation and the Limber's approximation and thus
  apply to all angular scales and redshift bins relevant. 

\bfi{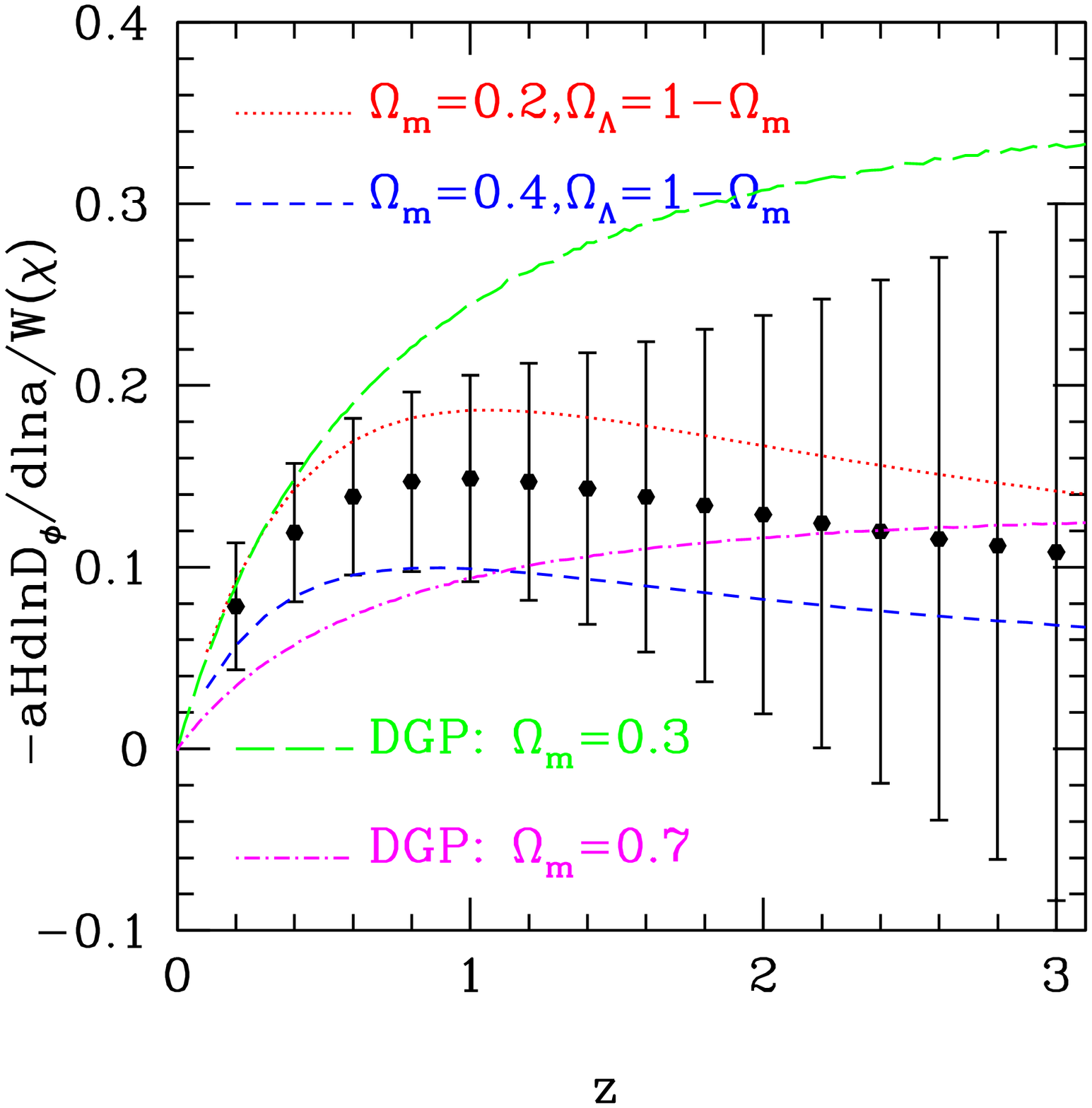}
\caption{The accuracy of measured $f\equiv aH d\ln D_{\phi}/d\ln
  a/W(\chi)$, assuming both CMB and galaxy surveys cover the full
  sky. Fiducial cosmology has
  $(\Omega_m,\Omega_{\Lambda},\Omega_b,h,\sigma_8)=(0.268,0.732,0.044,0.71,0.84)$. 
  Transfer function is calculated using the fitting formula of
  \citep{Eisenstein98}. 
\label{fig:Dphi}}
\efi

One can construct the projected gravitational potential $\Phi$ (or its
harmonic mode) from CMB
lensing \citep{Seljak99,Zaldarriaga99,Hu02}, 21cm background lensing \citep{Zahn05}, cosmic
shear (see \citet{Refregier03} for a recent review) and cosmic
magnification \citep{Zhang04}. $\Phi(\hat{n})=-\int d\chi
(\phi-\psi)W(\chi,\chi_s)$, where $W=(1-\chi/\chi_s)/\chi$, $\chi$ and 
  $\chi_s$ are the comoving angular diameter distance to lens and
  source. The cross correlation between $\Phi$ and LSS is
\be
\label{eqn:phig}
C_{\Phi g}=4\pi
\int
\Delta^2_{\phi\delta}(k,z=0)\frac{dk}{k}A_{\Phi}(k,l)A_g(k,l)
\ee
where $A_{\Phi}=2\int_0^{\chi_s}
j_l(k\chi)W(\chi,\chi_{s})D_{\phi}d\chi$, when $\phi+\psi=0$.

Eq. \ref{eqn:Ig} \& \ref{eqn:phig} can be further simplified by the Limber's
approximation \citep{Limber54,Kaiser98}. \cite{Afshordi04} showed that, even for a shallow survey such as
2MASS, which has an equivalent $\Delta z \sim 0.1$, the Limber's
approximation  is still valid to several percent level for
$l\geq 3$. We then adopt $\Delta z=0.2$ and apply the Limber's
approximation. Under this approximation, we have
\be
C_{Ig}=\frac{4\pi^2}{l^3}\int_{\chi_1}^{\chi_2}\Delta^2_{\phi\delta}(\frac{l}{\chi},z=0)n_gD_g
\dot{D}_{\phi}a\chi d\chi
\ee
Since $D_{\phi}$, $\dot{D}_{\phi}$, $a$, $\chi$,
$\Delta^2_{\phi\delta}(\frac{l}{\chi})$ do not vary significantly
over the redshift range $[\bar{z}-\Delta z/2,\bar{z}+\Delta z/2]$ as
long as that $\bar{z}\ga \Delta z$ and $n_g$ does not vary
significantly, we have  
\be
C_{Ig}\simeq \frac{4\pi^2}{l^3}[\Delta^2_{\phi\delta}(\frac{l}{\chi},z=0)n_gD_g
\dot{D}_{\phi}a\chi]|_{\bar{z}} (\chi_2-\chi_1)\ .
\ee
For the same reason, we have
\ba
C_{\Phi g} &\simeq& \frac{4\pi^2}{l^3}[\Delta^2_{\phi\delta}(\frac{l}{\chi},z=0)n_gD_g
D_{\phi}W\chi]|_{\bar{z}} (\chi_2-\chi_1)\ .
\ea
These approximations are accurate to several percent level for relevant $l$ and
  $\bar{z}\geq 0.2$.  We then have 
\be
C_{Ig}\simeq
f(\bar{z})C_{\Phi g}\ .
\ee Here, $f(z)=[(d\ln D_{\phi}/d\ln a)]aH/W(\chi)$, is the
  key quantity we want to measure.  Given the large cosmic variance in the $C_{Ig}$
  measurement,  one can safely neglect any possible errors caused by
  this approximation. Since $C_{Ig}\simeq
f(\bar{z})C_{\Phi g}$ is a key relation in this paper, here we show
another proof.  The Limber's approximation is achieved by taking the limit that
$j_l(x)\rightarrow \sqrt{\pi/(2l+1)}\delta_D(l+1/2-x)$. Under this
limit, for each $l$, only  those $k$ between
$[(l+1/2)/\chi_2,(l+1/2)/\chi_1]$ contribute to $C_{Ig}$, since only
those $k$ have non-vanishing $A_g(k,l)$. For these $k$,
$\int_0^{\chi_1}  
  j_l(k\chi)\dot{D}_{\phi}ad\chi\simeq 0$, $\int_{\chi_2}^{\chi_{\rm CMB}} 
  j_l(k\chi)\dot{D}_{\phi}ad\chi\simeq  0$ and $A_I(k,l)\simeq \int_{\chi_1}^{\chi_2} 
  j_l(k\chi)\dot{D}_{\phi}ad\chi$.  Similarly, we have
  $A_{\Phi}\simeq 2\int_{\chi_1}^{\chi_2}
j_l(k\chi)W(\chi,\chi_{s})D_{\phi}d\chi$ for these $k$. Thus, we have
$C_{Ig}\simeq 
f(\bar{z})C_{\Phi g}$, from Eq. \ref{eqn:Ig} \& \ref{eqn:phig}.

  $f(z)$ is determined by the matter
  density, dark energy density and equation of state, but it does not
  depend on galaxy bias or matter power spectrum. Since $H_0$,  the Hubble
  constant at present, in $H(z)$ and $W(\chi)$ cancel each other, $f$
  does not depend on $H_0$, too. $f(z)$ of each redshift bins can be estimated from the estimator $\hat{f}=\sum_l
C_{Ig}(l)w_l/\sum C_{\Phi g}w_l$, where $w_l$ is the weighting
function. For  $\Phi$ reconstructed from future cosmic shear, cosmic
magnification and CMB lensing,  the
fractional error in the denominator is small and one can do Taylor
expansion to estimate the error in $\hat{f}$. For this limit, we have
\be
\label{eqn:fullerror}
\frac{\Delta f^2}{f^2}=\frac{\sum_l\Delta C^2_{Ig}w_l^2}{[\sum_l
    C_{Ig}w_l]^2}+\frac{\sum_l\Delta C^2_{\Phi g}w_l^2}{[\sum_l
    C_{\Phi g}w_l]^2}-\frac{2\sum_l \langle \Delta C_{Ig}\Delta_{\Phi
    g}\rangle w_l^2}{\sum_l
    C_{Ig}w_l\sum_l
    C_{\Phi g}w_l} \ .
\ee
Here, 
\ba
\Delta
C^2_{Ig}(l)&=&\frac{\tilde{C}_{\rm CMB}\tilde{C}_g+C_{Ig}^2}{(2l+1)f_{\rm 
    sky}}\ , \nonumber \\
\Delta
C^2_{\Phi g}(l)&=&\frac{\tilde{C}_{\Phi}\tilde{C}_g+C_{\Phi g}^2}{(2l+1)f_{\rm 
    sky}} \ , \\
\langle \Delta C_{Ig}\Delta C_{\Phi g}\rangle&=& \frac{C_{\Phi
    I}\tilde{C}_g+C_{Ig}C_{\Phi g}}{(2l+1)f_{\rm 
    sky}} \nonumber
\ea
are the statistical errors of $C_{Ig}$, $C_{\Phi g}$ and cross
correlation between them, respectively. $\tilde{C}_{\rm CMB}$,
    $\tilde{C}_g$ and $\tilde{C}_{\Phi}$ are the sums of corresponding
    signals and associated contaminations. For $\Phi$ reconstructed from CMB
lensing of a very low noise CMB experiment,  $\tilde{C}_{\Phi}\simeq C_{\Phi}$ at $l<200$
\citep{Hu02}. For $\Phi$ reconstructed from the cosmic magnification
    and cosmic shear of future radio and optical surveys, 
this holds over even larger $l$ range. Since most ISW-LSS cross correlation signal
    comes from $l\la 100$ \citep{Afshordi04b} and 21cm emitting
    galaxies have high surface density,  we neglect shot noise
    term\footnote{For the estimations of  LSS clustering signal and
    shot noise, 
biggest uncertainties are (1) HI ({\it neutral hydrogen}) mass function at
high $z$, (2) 21cm emitting
galaxy bias and (3) specifications of 21cm experiments.  If one adopts HI mass
functions calibrated against observations of damped
Lyman-$\alpha$ systems and Lyman limit systems, SKA can detect $\sim  10^9$ galaxies
at $z\sim 3$ in five years across the whole sky, for a field of view $10$ deg$^2$ at $\sim 300$ Mhz (for
details of the calculation, see, e.g. \cite{Zhang06}). For SKA, detection
threshold of  HI mass at $z=3$ is $\sim 10^9M_{\odot}$, so detected
galaxies are likely having biases bigger
than one.  Then, one can neglect the shot noise term with respect to
    $C_g$} and approximate  $\tilde{C}_g= C_g$. Since
$C^{\rm CMB}/C^{\rm ISW}\gg 1$, errors in $C_{Ig}$ 
will dominate over errors in $C_{\phi g}$ and cross
correlations. Furthermore, the last two terms in Eq. \ref{eqn:fullerror}
    partly cancel. So it is a good approximation to neglect the last
    two terms. Under this simplification, we obtain the minimum variance
    estimator $w_l=\frac{C_{Ig}(l)}{\Delta C^2_{Ig}(l)}$ 
and the minimum variance is
\be
\frac{\Delta f^2}{f^2}\simeq \left(\sum_l \frac{C_{Ig}^2}{\Delta
  C_{Ig}^2}\right)^{-1}\simeq \left(\sum_l \frac{(2l+1)f_{\rm
  sky}C^{\rm ISW}}{r^2C^{\rm CMB}}\right)^{-1}
\ee
where $C^{\rm ISW}$ is the ISW power spectrum of the corresponding
    redshift bin and $r$ is the cross correlation coefficient between the ISW effect
    of the corresponding redshift bin
and galaxies. For narrow redshift bins we choose, $r$ is very close to unity.

 We choose the fiducial cosmology with best fit WMAP parameters
 $\Omega_m=0.268$, 
$\Omega_{\Lambda}=1-\Omega_m$, 
$\Omega_b=0.044$, $h=0.71$, $\sigma_8=0.84$ and  primordial power index
$n=1$. Transfer function is calculated according to
\citet{Eisenstein98}.  We use this fiducial cosmology to estimate the
 S/N of $f$ to be measured from future observations. We then use this
 fiducial  S/N to  forecast dark energy and gravity constraints. 
We choose 21cm emitting galaxies as
tracers of LSS. 21cm emission is less affected by dust and 21cm
surveys such as SKA can cover the whole sky. They also have the
 advantage to go deep into $z>3$ and allow the measurement of $f$ at $z\sim
 3$. We choose the CMB reference 
experiment discussed in \citet{Hu02} to produce a full sky lensing
map. Galaxy bin size is chosen to be $\Delta z=0.2$. For each galaxy
 redshift bin, one obtains a measure of 
 $f(z)$ at that redshift.  The result is 
shown in Fig. \ref{fig:Dphi}. For the fiducial model, the decay of
 $\phi$ can be detected at 
$z<2$. Since the physical separation of two redshift bins are much
larger than the galaxy correlation length, LSS of any two redshift
bins are uncorrelated, $f$ of each bins are close to be
independent\footnote{Because the ISW effect is  projected over the redshift
  range from zero to $\sim 1100$, 
  there does exist a correlation between $C_{Ig}(l)$ of different
  redshift bins. The same is true for $C_{\Phi g}$.  However,
  statistical errors of $f$ in each redshift bins are dominated by
  primary CMB. So the cross correlation coefficients between error bars
  of different redshift bins are close to zero. It is
  safe to neglect such correlations and take $f$ measurements of each
  redshift bins as uncorrelated.}.  We will do a reduced $\chi^2$ analysis to
estimate the dark energy constraints and gravity constraints. The reduced $\chi^2$
is defined as 
\be
\Delta \chi^2=\sum_i \frac{(f(z_i)-f^{\rm fid}_i)^2}{\Delta f^{\rm
    fid,2}_i} \ ,
\ee
where $f(z_i)$ and $f^{\rm fid}$ are the predicted $f$ of a given dark energy model or
gravity model and of the fiducial model, respectively. 
\bfi{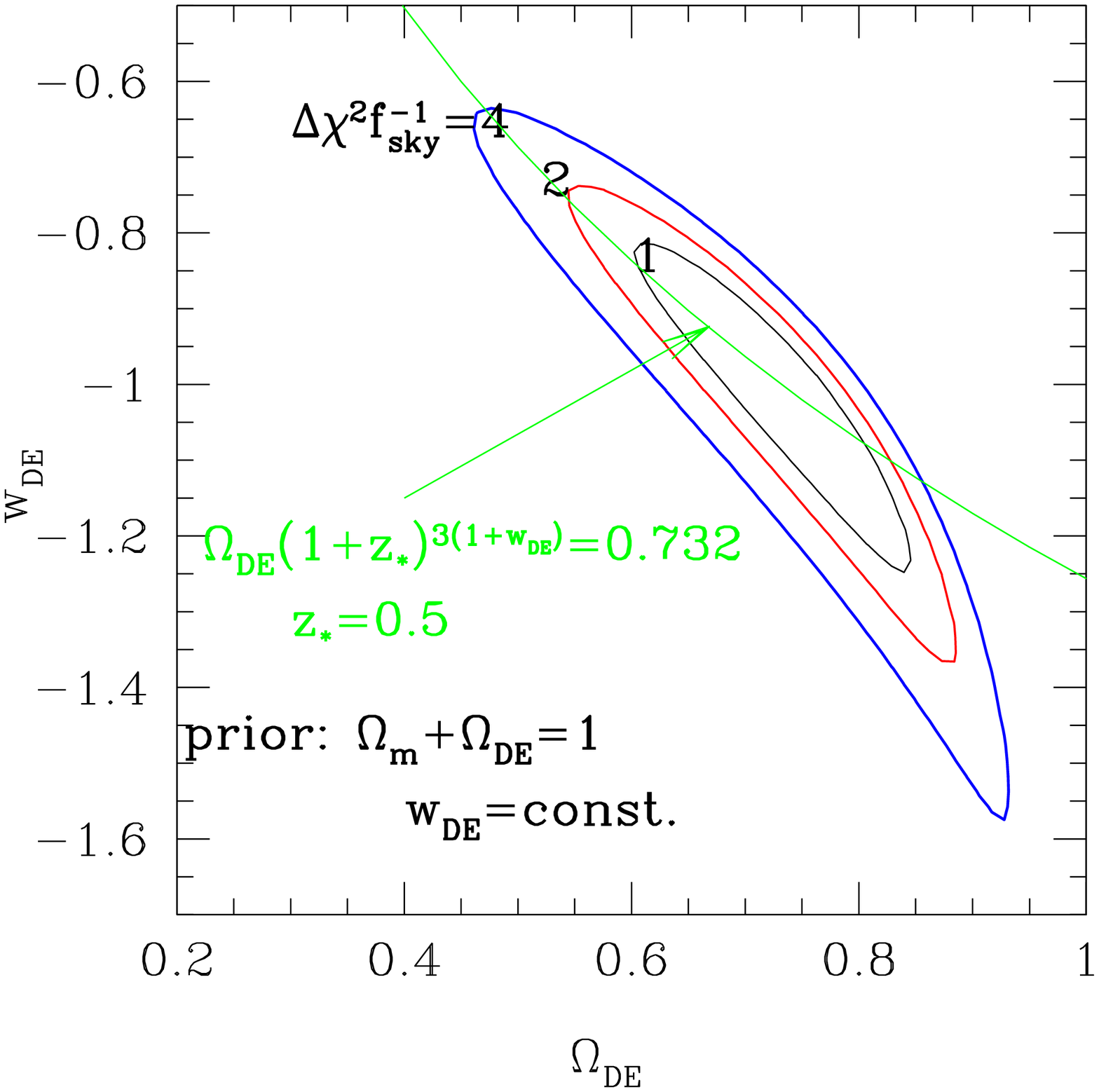}
\caption{$\Omega_{\rm DE}$-$w_{\rm DE}$ contour. Since most signals
  come from $z_*\sim 0.5$, $d\ln D_{\phi}/d\ln a$ at $z_*\sim 0.5$ should
  be close to the fiducial cosmology in order to get a good fit. This
  requires that the dark energy density $\Omega_{\rm
  DE}(1+z_*)^{3(1+w_{\rm DE})}\simeq \Omega_{\Lambda}^{\rm
  fiducial}$. We plot the line with $z_*=0.5$. \label{fig:DE}}
\efi

\section{Constraining dark energy}
The quantity $f$ is very sensitive to the amount of dark energy. To
visualize it, we plot those of $\Omega_m=0.2,0.4$ flat $\Lambda$CDM in
Fig. \ref{fig:Dphi}. One can see that these two cosmologies can be
distinguished from the fiducial cosmology at $>3\sigma$ confidence
level (C.L.). Given  a flat
$\Lambda$CDM prior,  $\Omega_{\Lambda}$ can be
constrained to $0.73^{+0.05}_{-0.07}$ at $2\sigma$ C.L.. 

We want to constrain both $\Omega_{\rm DE}$ and $w_{\rm DE}$
simultaneously. At super-horizon and near horizon scales, dark energy fluctuation can
contribute  $\sim 10\%$ to the decay of $\phi$ \citep{Bean04,Hu04}. To calculate the dark
energy fluctuation, one extra parameter, the sound speed $c_s$, is
required. Fortunately, the ISW signal (weighted by noise) mainly comes
from sub-horizon scale. For the purpose of this paper, we can neglect
the  dark  energy  
fluctuation in the parameter estimation. With this simplification,
$D_{\phi}$ is given by  
\be
D_{\phi}^{''}+D_{\phi}^{'}(\frac{5}{a}+\frac{H^{'}}{H})+\frac{D_{\phi}}{a^2}(3+\frac{H^{'}a}{H}-\frac{3}{2}\frac{\Omega_m  
  H_0^2}{a^3H^2})=0  \ ,
\ee
where $^{'}\equiv d/da$ and $^{''}=d^2/da^2$.
We numerically solve this equation with the initial condition that
when $a\rightarrow 0$, $D_{\phi}\rightarrow $const. and $D^{'}_{\phi}\rightarrow
0$. The results is shown in Fig. \ref{fig:DE}. The constraints are not very 
impressive, comparing to other methods. But since the theoretical
prediction only involves general relativity and parameter fitting
involves the least amount of free
parameters, the constraints are less affected by model uncertainties and the existence of dark
energy can be confirmed at high confidence level.

\section{Distinguishing dark energy from  modified gravity}
Precision
cosmology provides crucial tests of general relativity at  cosmological
scales. Modifications to general relativity change not only the
expansion history of the Universe, but also the LSS. For modified
gravities which reproduce the expansion history of dark energy
models (e.g. \citet{DGP,Carroll04,Carroll05,Mena05}),  the structure growth is in general
different. Thus, the measured $f$ from our methods can provide crucial
tests for such models. 
In general, modifications to general relativity causes $D_{\phi}$ to
  be scale dependent. In this case, one has to choose much narrower $l$
  bin to isolate $f(z)$ of each scale. Since the relative error in the
  denominator can approach unity for narrow $l$ bin, more
  sophisticated estimator is required to isolate $f$.  For simplicity,
  in this paper we only discuss the
  Dvali-Gabadadze-Porrati (DGP) model \citep{DGP}, which preserves
  the scale independent $D_{\phi-\psi}$ \citep{Lue04b,Koyama05}.  

\bfi{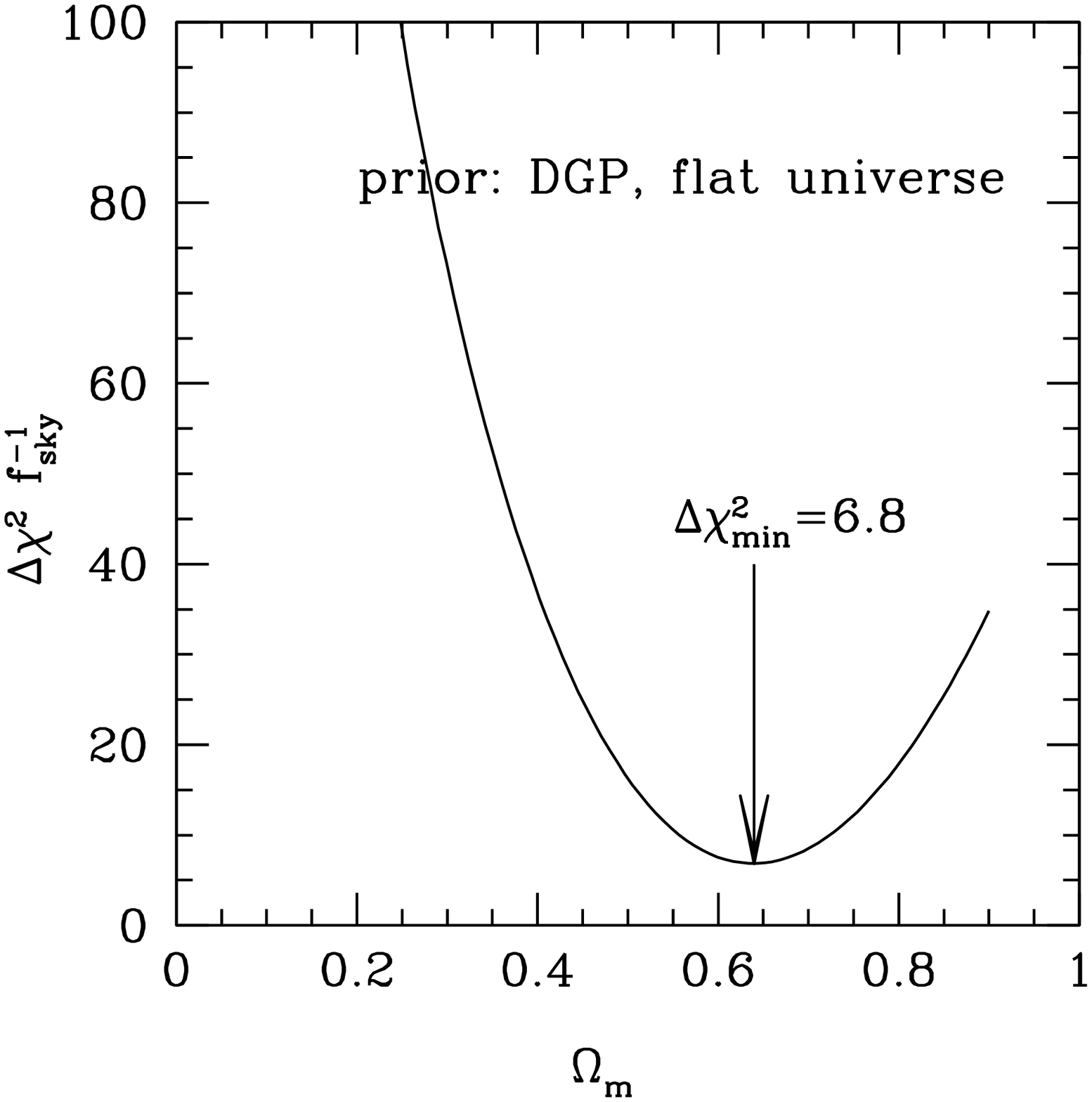}
\caption{Distinguishing DGP from the fiducial $\Lambda$CDM
  cosmology. Since at high redshifts gravitational potential decays much faster in DGP
  than in $\Lambda$CDM cosmology, DGP can
  be distinguished from $\Lambda$CDM at $>2.5\sigma$ confidence level. }
\efi
For a flat DGP model, the expansion history is given by
  $H^2=\frac{H}{r_0}+H_0^2\Omega_ma^{-3}$, 
where $r_0=1/[H_0(1-\Omega_m)]$.  The two Newtonian potential
$\phi$ and $\psi$ no longer follow the relation $\phi+\psi=0$. The
  evolution of  $\phi-\psi$ is governed by
\be
D_{\phi-\psi}^{''}+D_{\phi-\psi}^{'}(\frac{5}{a}+\frac{H^{'}}{H})+\frac{D_{\phi-\psi}}{a^2}(3+\frac{H^{'}a}{H}-\frac{3}{2}\frac{\Omega_m  
  H_0^2}{a^3H^2}\frac{G_{\rm eff}}{G})=0 
\ee
where $G_{\rm
  eff}/G=1+1/3\beta$ and $\beta=1+2r_0H^2/(1-2r_0H)<0$
  \citep{Lue04b,Koyama05}. This equation holds at sub-horizon scales
  with $k\gg aH$ and $k\gg 1/r_0$. At $z\ga 2$, $l<10$ modes do not
  satisfy these conditions. But at these redshifts, most signal comes
  from $l>10$ and throwing away $l<10$ modes does not degrade the $f$
  measurement significantly. For simplicity, we neglect this
  complexity. 

We find that flat DGP model can be  
distinguished from the fiducial $\Lambda$CDM cosmology at $>2.5\sigma$ C.L.. Fig. 1
demonstrates this point. When $a\rightarrow 0$, the expansion history
of a DGP model resembles a $w_{\rm DE}=-1/2$ dark energy model (or
$\Lambda$CDM with curvature). But gravitational  
  potential in  DGP decays faster,  due to extra suppression caused by smaller effective
  Newton's constant $G_{\rm eff}$.  Comparing to the fiducial
  $\Lambda$CDM model, a low $\Omega_m$ DGP model causes
  too fast gravitational potential decay at high redshifts while a high $\Omega_m$ DGP
  model causes too slow gravitational potential decay at low
  redshifts.  So it is hard for DGP model to reproduce $f$ and its
  redshift dependence in  
  the fiducial $\Lambda$CDM cosmology. 
One can infer similar conclusion From Fig. 3, where $w_{\rm DE}>-0.6$
dark energy model is excluded at $> 90\%$ confidence level. Since
gravitational potential in DGP decays faster than a $w_{\rm DE}=-1/2$ dark
energy model, DGP should be distinguished from the fiducial
model with higher C.L.. This can be 
better understood from the asymptotic behavior of $\dot{D}_{\phi}$. When $a\rightarrow 0$, 
$d\ln
D_{\phi-\psi}/d\ln a\rightarrow
-[11/16][(1-\Omega_m)/\Omega_m^{1/2}]a^{3/2}$ for DGP and
$d\ln D_{\phi-\psi}/d\ln a\rightarrow -[3(1-w_{\rm DE})/(6-5w_{\rm
    DE})][(1-\Omega_m)/\Omega_m]a^{-3w_{\rm DE}}$ for dark energy.  At
  high redshifts, gravitational potential decays  faster in DGP than in
  dark energy models with $w_{\rm DE}\leq 
  -1/2$ and much faster than in $\Lambda$CDM cosmology. It is this
  feature that allows a clear discrimination between  
DGP and many dark energy models  using the ISW effect. 

\section{Discussion}
We have shown how to isolate the decay rate of
gravitational potential from ISW-LSS and lensing-LSS cross correlation
measurements. The measured decay rate, with prefactors only depend on
geometry of the universe can put robust constraints on dark energy and
the nature of gravity, free of many theoretical uncertainties. The
accuracy of such measurement is dominated 
by statistical fluctuations dominated by primacy CMB. This allows us
to safely neglect several possible systematics. (1) Dark energy
fluctuations. As shown in \citet{Bean04}, such 
  fluctuations are at most  $10\%$ those of dark matter
  fluctuation. For sound speed close to unity, these fluctuations
  effectively vanishes at most scales we are interested and the
  overall effect can be neglected. (2) Time evolution of $f(z)$. Since
  the bin size $\Delta z$ is not infinitely 
  small, the simplification $f(z)\simeq f(\bar{z})$ for
  $\bar{z}-\Delta z/2<z<\bar{z}+\Delta z/2$ causes an relative error $
  \simeq [d^2f/dz^2/f][(\Delta z)^2/12]\ll 1$, for adopted $\Delta
  z=0.2$. (3) Correlated error bars in $f$ measurement.  

We keep in caution that foregrounds of CMB, LSS and lensing may be correlated and their
  effects may become non-negligible toward the 
  galactic plane. In this case, one needs to mask the galactic plane, at the 
  expense of losing statistical accuracy, scaled as $f_{\rm sky}$ (fractional sky coverage).

{\it Acknowledgments}.--- The author is supported  by the
One-Hundred-Talent Program of China and the NSFC grant
(No. 10533030). The author thanks Fermilab theoretical  
astrophysics group for the hospitality where this work was
finalized. The author thanks the anonymous referee for helpful
suggestions.

\end{document}